\documentclass{article}
\usepackage{spconf,amsmath,graphicx,url,times,amsmath,amssymb,acronym,graphicx,balance}
\pdfoutput=1
\usepackage{color,booktabs}
\usepackage[table]{xcolor}
\usepackage{cite}
\usepackage{lipsum}

\title{A consolidated view of loss functions for supervised deep learning-based speech enhancement}
\name{Sebastian Braun, Ivan Tashev} 
\address{
Microsoft Research, Redmond, WA, USA \\
sebraun@microsoft.com, ivantash@microsoft.com}

\acrodef{STFT}{short-time Fourier transform}
\acrodef{MSE}{mean-squared error}
\acrodef{MAE}{mean absolute error}
\acrodef{PSD}{power spectral density}
\acrodef{RTF}{relative transfer function}
\acrodef{SNR}{signal-to-noise ratio}
\acrodef{PDF}{probability density function}
\acrodef{DOA}{direction-of-arrival}
\acrodef{VAD}{voice activity detector}
\acrodef{MVDR}{minimum variance distortionless response}
\acrodef{AIR}{acoustic impulse response}
\acrodef{PESQ}{perceptual evaluation of speech quality}
\acrodef{STOI}{short-time objective intelligibility}
\acrodef{LSD}{log spectral distance}
\acrodef{CD}{cepstral distance}
\acrodef{WER}{word error rate}
\acrodef{SPP}{speech presence probability}
\acrodef{RNN}{recurent neural network}
\acrodef{LSTM}{long-term short-term}
\acrodef{GRU}{gated recurrent unit}
\acrodef{FF}{feed forward}
\acrodef{ReLU}{rectified linear unit}
\acrodef{GCC}{generalized cross-correlation}
\acrodef{RMSE}{root-mean-square error}
\acrodef{CPSD}{cross-power spectral density}
\acrodef{SI-SDR}{scale-invariant signal-to-distortion ratio}
\acrodef{SDR}{signal-to-distortion ratio}
\acrodef{MagMSE}{Magnitude MSE}
\acrodef{LPS}{logarithmic power spectrum}
\acrodef{CSE}{complex spectrum error}
\acrodef{MagSE}{magnitude spectrum error}
\acrodef{LogMagSE}{logarithmic magnitude spectrum error}
\acrodef{DL}{deep learning}
\acrodef{PLSD}{phase-aware logarithmic spectrum distance }
\acrodef{SDW}{speech distortion-weighted}

\providecommand{\numberTblEq}[1]{\refstepcounter{tblEqCounter}\label{#1}\thetag{\thetblEqCounter}}

\begin{document}
\ninept

\maketitle
\begin{abstract}
Deep learning-based speech enhancement for real-time applications recently made large advancements. Due to the lack of a tractable perceptual optimization target, many myths around training losses emerged, whereas the contribution to success of the loss functions in many cases has not been investigated isolated from other factors such as network architecture, features, or training procedures. 
In this work, we investigate a wide variety of loss spectral functions for a recurrent neural network architecture suitable to operate in online frame-by-frame processing. We relate magnitude-only with phase-aware losses, ratios, correlation metrics, and compressed metrics. Our results reveal that combining magnitude-only with phase-aware objectives always leads to improvements, even when the phase is not enhanced. Furthermore, using compressed spectral values also yields a significant improvement. On the other hand, phase-sensitive improvement is best achieved by linear domain losses such as mean absolute error.
\end{abstract}


\newcounter{tblEqCounter} 

\section{Introduction}
\label{sec:intro}
Speech enhancement using neural networks has seen large attention and success in the recent years \cite{Wang2018}. While classic single-channel statistical model-driven speech enhancement techniques used in practical systems often only leverage signal models for quasi-stationary noise \cite{Kavalekalam2018}, neural networks can potentially learn more complex speech characteristics, which also allows reduction of highly non-stationary, transient noise, and non-speech sound sources.

Unfortunately, state-of-the-art \ac{DL} based noise reduction performance is currently only achieved by 
architectures requiring large look-ahead, large amounts of temporal context data input \cite{Williamson2017,Ephrat2018,Luo2019}, or computationally expensive network architectures \cite{Williamson2017,Strake2019,Wichern2017,Wisdom2019,Tan2018}. As the performance seems to scale with the network size this often prohibits the use in real-time speech communication systems such as live-messengers or mobile communication devices.


However, the training loss function is independent of the inference complexity, and has therefore potential to improve performance at no cost.
Although the most popular choice for regression-based \ac{DL} is the \ac{MSE}, this might arguably be not the optimal choice for speech enhancement. Loss functions and training targets for speech enhancement have shifted from the \ac{MSE} between several versions of enhancement filters or masks \cite{Wang2014,Williamson2017} to signal-based metrics, such as spectral magnitude-based \ac{MSE}, phase-sensitive \ac{MSE} \cite{Kolbaek2017} and finally the complex spectral \ac{MSE} \cite{Strake2019}.
Approaches originating from a source separation background often use the time-domain \ac{MSE} or \ac{SDR} loss \cite{Luo2019,Bahmaninezhad2019}.

While recent attempts were made integrating perceptually motivated metrics in the loss function \cite{Martin-Donas2018,Zhao2019}, optimizing on perceptual metrics alone is often insufficient, and is therefore combined again with lower-level criteria such as the spectral magnitude \ac{MSE}.
It is often observed that optimization on some objective metrics like \ac{PESQ} or \ac{STOI} improves the test results for the optimized metric, but fails to outperform other baselines in terms of other metrics \cite{Martin-Donas2018,Zhao2019}. 
While the log-energy sigmoid weighting proposed in \cite{Liu2017} does not generalize as it is highly heuristic and signal level dependent, we also could not verify improvements using a noise shaping weighting as proposed in \cite{Zhao2019} for our tested networks and data. 
Therefore, we take a step back and investigate different basic signal distance metrics as optimization criteria, which does not exclude the possibility to add perceptually motivated weightings.

As in the last years a large variety of speech enhancement loss functions have been proposed, it is impossible to quantify their individual contribution to success due to the use of different enhancement systems and datasets.
The study in \cite{Kolbaek2020} compares a selection of loss functions for a convolutional time-domain network. These results may differ greatly from our study due to a complex network architecture with larger delay, an inference complexity more than 30 times larger than our network, and training/evaluation on non-reverberant speech, which is rarely encountered in practice.
In this work, we compel an overview and comparison of different frequency-domain optimization criteria using a small recurrent neural network suitable for on-the-edge real-time inference. We classify the losses based on their distance metric in spectral magnitude and complex losses, propose some new losses closing gaps in this systematic search, and point out interesting relations.
We show that the best performing of the tested loss functions are the compressed \ac{MSE}, closely followed by the \ac{MAE}, which can be attributed to a better match to the signal distributions. We furthermore show that linear combination of magnitude and complex losses leads to improvement in all cases.
Another interesting finding is that our results on a reverberant speech dataset did not confirm advantages of the recently proposed \ac{SDW} \cite{Xia2020} and noise shaping losses \cite{Zhao2019}.



\section{Signal Model}
\label{sec:sigmodel}
In a pure noise reduction task, we assume that the observed signal is an additive mixture of the desired speech and noise. We denote the observed signal $X(k,n)$ directly in the \ac{STFT} domain, where $k$ and $n$ are the frequency and time frame indices as
\begin{equation}
X(k,n) = S(k,n) + N(k,n),
\end{equation}
where $S(k,n)$ is the potentially reverberant speech, and $N(k,n)$ is the disturbing noise signal. 
%
The objective is to recover a speech signal estimate $\widehat S(k,n)$ by
\begin{equation}
\label{eq:Shat}
\widehat S(k,n) = G(k,n) \, X(k,n).
\end{equation}
where $G(k,n)$ is a filter that can be either a real-valued suppression gain, or a complex-valued filter. In this work, we consider only a suppression gain.

\section{Loss functions}
\label{sec:lossfunctions}
In this section, we review and introduce a wide range of training loss functions targeting recovery of the speech signal $S(k,n)$. 
%
All considered speech enhancement loss functions are distance metrics between the enhanced and target spectral representations. 
We can classify the loss functions summarized in Table~\ref{tab:lossfuns} in magnitude distances and complex spectral distances, which also incorporate phase information. The operator $\left<Y(k,n)\right> = \frac{1}{KN} \sum_{k,n} Y(k,n)$ denotes the arithmetic average over frequency and time indices, $k$ and $n$, per sequence. Newly proposed loss functions are marked with a $\dagger$.
In the following, we introduce and discuss the loss functions in Table~\ref{tab:lossfuns}.

\setcounter{tblEqCounter}{\theequation}
\rowcolors{2}{gray!25}{white}
\begin{table*}[tb]
	\caption{Loss functions}
	\label{tab:lossfuns}
	\small
	\centering
	\renewcommand{\arraystretch}{2}
	\begin{tabular}{c|lcc|lcc}
		\rowcolor{gray!50}
		metric 	&  	& magnitude 	& &  & complex 	& \\\toprule
		L2 		& magMSE & $\left<|\widehat{A}-A|^2\right>$ & \numberTblEq{eq:mse2} 	& 
		cMSE & $\left<|\widehat{S}-S|^2\right>$ & \numberTblEq{eq:cse2}  \\
		L1 		& magMAE & 	$\left<|\widehat{A}-A|\right>$ & \numberTblEq{eq:mse1}	& 
		cMAE & $\left<|\widehat{S}-S|\right>$ & \numberTblEq{eq:cse1} \\
		log MSE & LSD & $\left<|\log_{10} \widehat{A} - \log_{10}A|^2\right>$ & \numberTblEq{eq:LSA} 	
		& \multicolumn{2}{l}{PLSD$^\dagger$ \; $\left< |\log_{10} |\frac{\widehat{S}}{S}| \times \left(2 - \mathcal{R}\{\frac{\widehat{S}}{S}\} \right) \right>$} & \numberTblEq{eq:PLSD} \\
		weighted log MSE & wLSD$^\dagger$ & $ \left< W_\text{lsd} \, |\log_{10} \widehat{A} - \log_{10}A|^2 \right>$ & \numberTblEq{eq:wLSD} 	& wPLSD$^\dagger$ & $\left< W_\text{lsd} \, \; (\texttt{-"-}) \; \right>$	& \numberTblEq{eq:wPLSD} \\
		compressed	& magComp & $\left<|\widehat{A}^c -A^c|^2\right>$ & \numberTblEq{eq:compmag}	&
		cComp & $\left<| \widehat{A}^c e^{j\varphi_{\widehat{s}}} - A^c	e^{j\varphi_{s}}|^2 \right>$ & \numberTblEq{eq:ccomp} \\
		ratios	& SNR$^\dagger$ & $-\log_{10} \frac{\left<A^2 \right>}{\left<|\widehat{A} - A|^2 \right>}$ & \numberTblEq{eq:snr} 	& 
		SDR & $-\log_{10} \frac{\left<|S|^2 \right>}{\left<|\widehat{S} - S|^2 \right>}$ & \numberTblEq{eq:sdr} \\
		correlation	& magCorr$^\dagger$ & $-\frac{\left<\widehat{A} A \right>^2}{\left<\widehat{A}^2\right> \left<A^2 \right>}$ & \numberTblEq{eq:xcorr_mag}	& 
		cCorr$^\dagger$ & $-\frac{\Re\left\{ \left< \widehat{S} S^* \right> \right\} }{\sqrt{\left< |\widehat{S}|^2\right> \left<|S|^2 \right>}}$  & \numberTblEq{eq:xcorr} \\
		speech dist. weight	& \multicolumn{2}{l}{SDW \; $\lambda \left<|S \!-\! GS|^2\right> + (1 \!-\! \lambda) \left<|GN|^2\right>$} & \numberTblEq{eq:sdnr} & & -- & \\
		weighted L2	& MSE-AMR & $\left<W_\text{AMR} \, |\widehat{A}-A|^2\right>$	&  \numberTblEq{eq:mse-amr} & cMSE-AMR & $\left<W_\text{AMR} \, |\widehat{S}-S|^2\right>$ &  \numberTblEq{eq:cmse-amr} \\
	\end{tabular}
\end{table*}
\setcounter{equation}{\thetblEqCounter}

\subsection{Linear spectral distance norms}
The most straightforward choice is the L2-norm or squared error between estimated and target signals. While this loss is often only magnitude based as in \eqref{eq:mse2} \cite{Weninger2015,Zhao2018a}, its complex counterpart \eqref{eq:cse2} is usually only used in direct spectral mapping approaches \cite{Fu2017,Strake2019}, but has strangely never been used in filter prediction networks so far.

An actually better distance metric for the complex error is the L1-norm or \ac{MAE}, as the distribution of \ac{STFT} bins follow a more Laplacian distribution rather than Gaussian, as can be observed in Fig.~\ref{fig:sig_distros} by the blue curves. The L1-norm of the magnitude and complex signal error are given by \eqref{eq:mse1} and \eqref{eq:cse1}, respectively, where we define the L1 norm of a complex number as $\Vert x_R + j x_I \Vert_1 = |x_R| + |x_I|$. The complex L1-norm loss \eqref{eq:cse1} has been termed \emph{RI} loss in \cite{Wang2020}.

%

\subsection{Logarithmic spectral distance}
\begin{figure}[tb]
	\centering
	\includegraphics[width=.9\columnwidth,clip,trim=50 0 90 10]{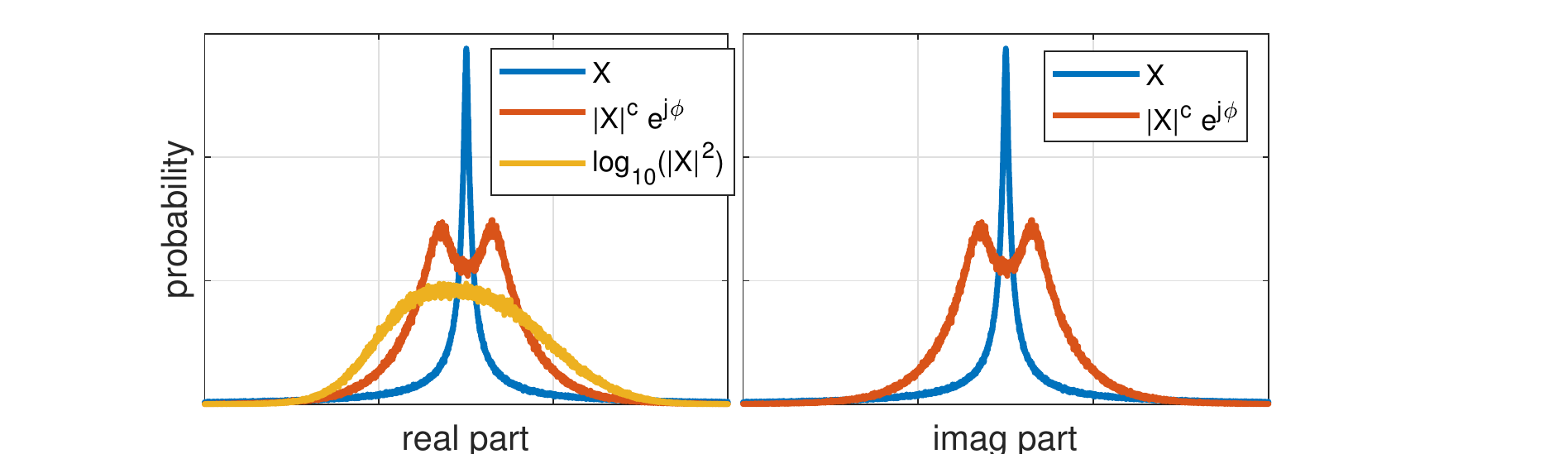}
	\caption{Distributions of linear complex, compressed complex and log spectral signals for 5~min noisy speech.}
	\label{fig:sig_distros}
\end{figure}

To account for the logarithmic perceptual nature of the human ear, the \ac{LSD} given by \eqref{eq:LSA} can be used, which was a standard in traditional model-based speech enhancement for decades \cite{Ephraim1985}.
Note that so far, the \ac{LSD} has only been proposed in methods directly predicting the log power spectrum instead of a filter \cite{Tu2018,Martin-Donas2018}, while we use it to predict a filter. The log compression creates a Gaussian-like distribution as shown in Fig.~\ref{fig:sig_distros} by the yellow line.

To extend the \ac{LSD} \eqref{eq:LSA} with a phase-error term, we propose the \ac{PLSD} by \eqref{eq:PLSD}, 
where $\varphi_S$ and $\varphi_{\widehat{S}}$ are the phase angles of $S(k,n)$ and $\widehat{S}(k,n)$, respectively. The first term in \eqref{eq:PLSD}, the magnitude error, is identical to \eqref{eq:LSA}. The second term, the phase error, is connected to the magnitude error by bin-wise multiplication, which naturally decreases the phase error at bins with small magnitude error. The constant 2 ensures that the phase error lies within the range of $[1,3]$, preventing vanishing magnitude error at zero phase error. Note that the cosine phase difference can be calculated as the real part of the complex signal division, i.e. $\cos(\varphi_{\widehat{S}} - \varphi_{S}) = \Re\left\{ \frac{\widehat{S}}{S}\right\}$.

\subsection{Weighted logarithmic loss}
Due to the logarithmic compression, the standard \ac{LSD} suffers from the problem of producing large errors also at low energy bins, which are perceptually less relevant. As limiting the log mitigates this problem only suboptimally, we propose to apply a bin-wise weighting based on the target speech signal in \eqref{eq:wLSD} with
\begin{equation}
\label{eq:LSDweight}
W_\text{wlsd}(k,n) = |\widehat{S}(k,n) + \gamma X(k,n)|^{0.3},
\end{equation}
where we chose $\gamma=0.1$ to blend in the noisy signal to prevent applying zero weights where high noise reduction is achieved, and apply a compression exponent of 0.3.
The same weighting can also be applied to the \ac{PLSD} as given by \eqref{eq:wPLSD}.

\subsection{Power-law compressed spectral distance}
A similar dynamic compression as the logarithm can be achieved using power-law compression \cite{Lee2018} applied to the magnitudes by \eqref{eq:compmag} with a compression exponent $0<c<1$. 

A phase-aware compressed loss can be obtained by multiplying the phase terms to the compressed magnitudes as given by \eqref{eq:ccomp}, which was proposed in \cite{Ephrat2018,Wilson2018}. A commonly used compression exponent is $c=0.3$. In contrast to the logarithm, this compression has the advantage of producing positive semi-definite values. We can observe in Fig.~\ref{fig:sig_distros} by the red lines that the compression broadens the distributions complex compressed spectra, while values closer to zero occur less frequent.


%

\subsection{Signal ratio losses}
Commonly used ratios in speech enhancement are the \ac{SNR} and \ac{SDR}. The time-domain \ac{SDR} has already been successfully used in \ac{DL} based speech enhancement \cite{Vincent2006a,Choi2019,Roux2019}. However, this metric is not restricted to the time-domain, and can be equivalently computed in the frequency domain.
We employ here the scale-variant \ac{SDR} given by \eqref{eq:sdr}, as we believe a scaled output signal as in the scale-invariant \ac{SDR} \cite{Roux2019} is undesired. 

In analogy, computing this ratio from magnitudes is more commonly termed the \ac{SNR}, given by \eqref{eq:snr}.
Note that the \ac{SNR} and \ac{SDR} losses, \eqref{eq:snr} and \eqref{eq:sdr}, are simply related to the \ac{MSE} losses \eqref{eq:mse2} and \eqref{eq:cse2}, normalized by the speech power, as was also pointed out in \cite{Heitkaemper2020}.

%

\subsection{Correlation based losses}
The speech intelligibility index and related objective metrics \cite{Taal2011a} are based on signal envelope correlation. Motivated by this fact, we introduce the magnitude correlation loss given by \eqref{eq:xcorr_mag}.

The complex equivalent, the complex correlation coefficient given by \eqref{eq:xcorr}, is better known as the \emph{coherence}.
While the range of \eqref{eq:xcorr_mag} is $[0,1]$, the range of \eqref{eq:xcorr} is $[-1,1]$.
%
%
Note that in \cite{Venkataramani2017}, the coherence loss \eqref{eq:xcorr} has been termed source-to-distortion ratio.
Special properties of the ratio and correlation based losses is that they are signal-level independent.

\subsection{Speech distortion weighted loss}
By using the signal components of speech and noise separately, the \ac{SDW} loss \cite{Xia2020,Xu2020} given by \eqref{eq:sdnr} provides a trade-off parameter $0<\lambda<1$ between speech distortion and noise reduction.
Note that while \eqref{eq:sdnr} does not explicitly use only magnitudes, the decomposed nature and absence of the noisy signal $X(k,n)$ implies that $G(k,n)$ as zero-phase filter is optimal. Therefore, the loss is categorized as magnitude loss.
Drawbacks of the \ac{SDW} loss are that the optimal weight $\lambda$ is data dependent, and finding optimal adjustments of $\lambda$ e.g. depending on the \ac{SNR}, are heuristic and difficult to determine.

\subsection{Weighted and combined losses}
In \cite{Zhao2019}, a weighting for the \ac{MSE} based on the AMR codec 
is proposed to spectrally shape the noise error. We include this loss given by \eqref{eq:mse-amr} and \eqref{eq:cmse-amr}, while also other weightings can be applied to most distance metrics.	

Several works have proposed combined losses using linear combinations of magnitude-only and phase-aware metrics \cite{Ephrat2018,Wisdom2019,Lee2018} as
\begin{equation}
\label{eq:mix_loss}
\mathcal{L}_\text{mix} = (1-\beta) \mathcal{L}_\text{mag} + \beta \mathcal{L}_\text{complex},
\end{equation}
where $\mathcal{L}_\text{mag}$ is a magnitude-based loss, $\mathcal{L}_\text{complex}$ is a complex signal based loss, and $0 \leq \beta \leq 1$ is the mixing factor. We investigate all useful combinations per row in Table~\ref{tab:lossfuns}.

%


\section{Network and training}
We use a recurrent network architecture based on \acp{GRU} \cite{Cho2014} and \ac{FF} layers, similar to the core architecture of \cite{Wisdom2019}, to estimate the enhancement filter $G(k,n)$. The architecture was chosen to maintain real-time constraints without delay and moderate complexity.

\begin{figure}
	\includegraphics[width=\columnwidth,clip,trim=100 140 70 120]{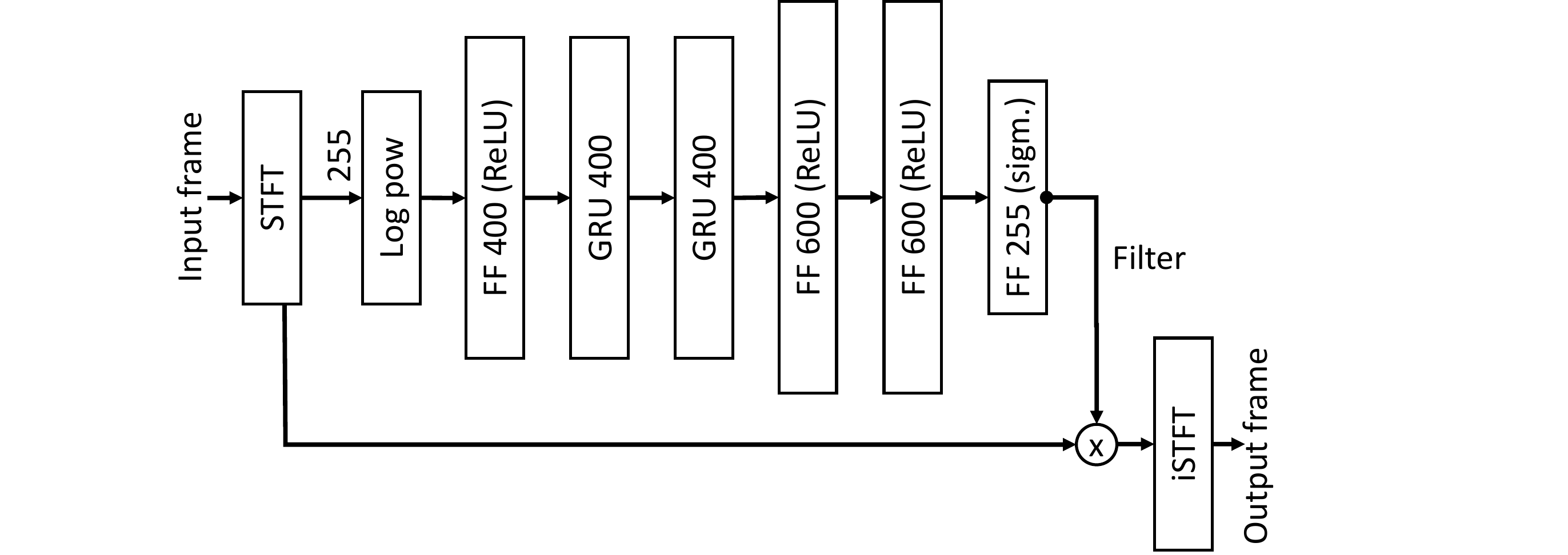}
	\caption{Network architecture and enhancement system.}
	\label{fig:architecture}
\end{figure}

The network input is the logarithmic power spectrum $P = \log_{10}(|X(k,n)|^2 + \epsilon)$ with online mean and variance normalization \cite{Xia2020}.
We use a \ac{STFT} size of 512 with 32~ms square-root Hann windows and 16~ms frame shift, but feed only the relevant 255 frequency bins into the network, omitting 0th and highest (Nyquist) bins, which do not carry useful information.
The network consists of a \ac{FF} embedding layer, two \acp{GRU}, and three \ac{FF} layers with \ac{ReLU} activations and an output layer with \emph{Sigmoid} activation.
The enhancement system and network architecture with layer sizes is shown in Fig.~\ref{fig:architecture}, and has 2.8~M trainable parameters.

The network was trained using the AdamW optimizer \cite{Loshchilov2019} with a learning rate of $10^{-4}$.
The training was monitored every 10 epochs using a validation subset. The best model was chosen based on the highest \ac{PESQ} \cite{ITU_T_P862} on the validation set.
Also the optimal weighting factors for $\beta,\lambda$ etc. were optimized by a grid search and choosing the best performing parameter for \ac{PESQ} on the validation set.

\section{Experiments}

\subsection{Dataset and evaluation metrics}
We used the Chime-2 WSJ-20k dataset \cite{Vincent2013}, which is, despite only of medium size, a realistic self-contained public dataset including matching reverberant speech and noise conditions. The dataset contains 7138, 2418, and 1998 utterances for training, validation and testing, respectively. The target speech signals are binaural and reverberant, and the mixtures contain noise recorded in the same rooms. Validation and test sets are mixed with \acp{SNR} in from -6 to 9~dB. For testing, we used only the left channel.
We evaluate the speech enhancement performance in terms of \ac{PESQ} \cite{ITU_T_P862} as an indicator for noise reduction and speech quality, and \ac{SI-SDR} \cite{Roux2019} as a phase-sensitive metric. 

\subsection{Results and discussion}
\begin{figure}[tb]
	\includegraphics[width=\columnwidth]{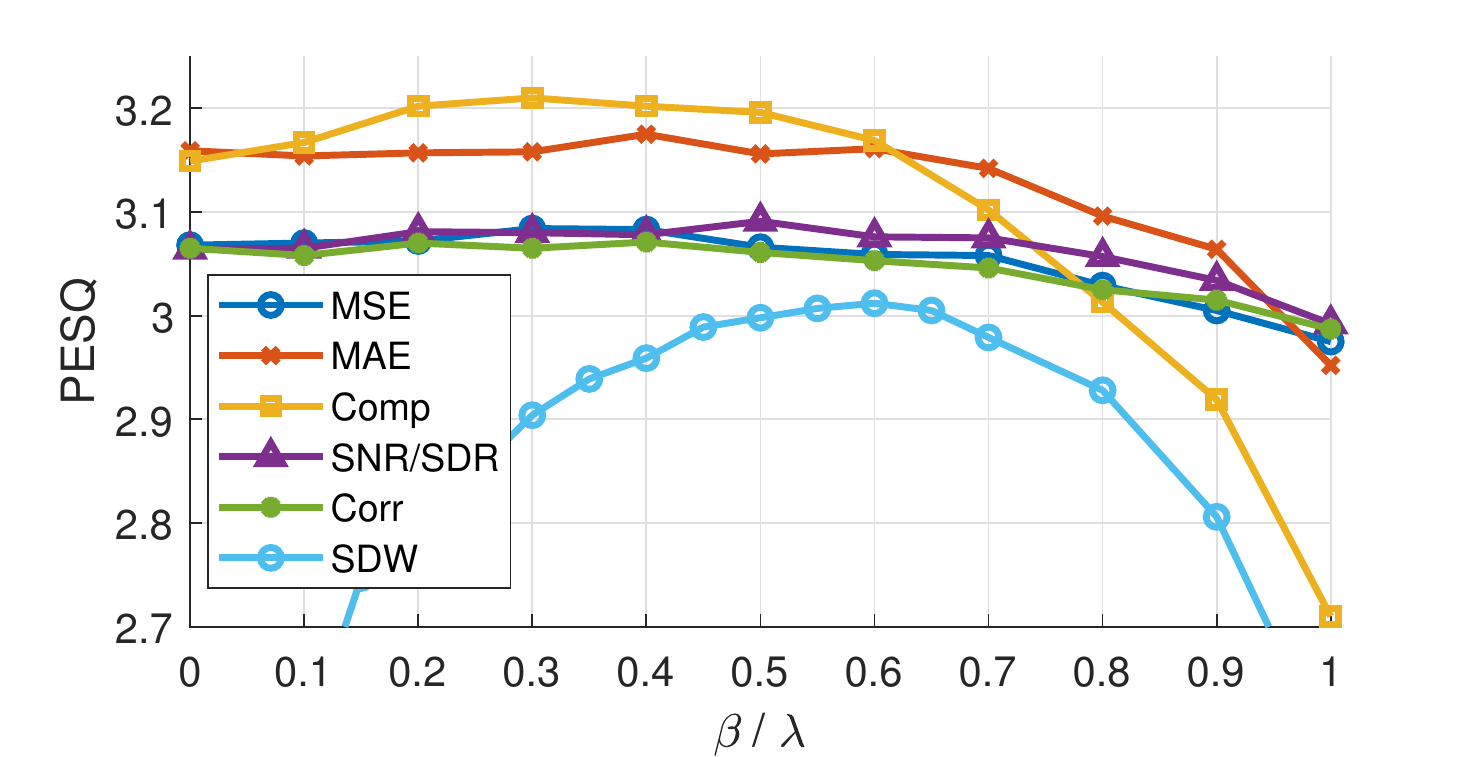}
	\caption{Optimization of magnitude vs. complex loss weight $\beta$ and speech distortion weight $\lambda$ on validation set.}
	\label{fig:complexweight}
\end{figure}
Each magnitude and complex loss per row in Table~\ref{tab:lossfuns} was combined by linear mixing \eqref{eq:mix_loss}. The \ac{LSD} losses were omitted as the PLSD is already a combined metric. The mixing factors were determined on the development set. The \ac{PESQ} results for the parameter sweeps of $\beta$ are shown in Fig.~\ref{fig:complexweight}. 
We can observe that the combination of magnitude and complex loss leads to an improvement for all distance metrics. We can also see that the \ac{MAE} and compressed losses outperform the other distance metrics significantly at the optimal weight $\beta$. Furthermore it is interesting, that the combined compressed loss of \eqref{eq:compmag}, \eqref{eq:ccomp} achieves the highest performance with $\beta=0.3$, while for magnitude loss only ($\beta=0$), compressed and \ac{MAE} are similar, but for fully complex loss ($\beta=1$), the compressed loss shows a significant performance drop.
Although we experimented with "out-of-metric" combinations, in particular combining magComp with a better complex loss, e.g. cMAE, this did not lead to an improvement.

The \ac{PESQ} and \ac{SI-SDR} results for all losses on the test set are shown in Table~\ref{tab:results}, where the combined losses in the right column use the \ac{PESQ}-optimal weightings. The best performers are highlighted in bold font. The \ac{PESQ} results align well with the development set in Fig.~\ref{fig:complexweight}, namely that \ac{MAE} and compressed loss are good performers. While the pure \ac{LSD} is even slightly worse than the \ac{MSE}, the signal power-weighted wLSD outperforms the linear \ac{MSE}.
\begin{table}[tb]
	\caption{PESQ (SI-SDR) on test set.}
	\label{tab:results}
	\centering
	\begin{tabular}{l|c c c c}
		\rowcolor{gray!50}
		loss	& magnitude		& complex	& comb. \eqref{eq:mix_loss} \\\toprule
		\rowcolor{white}
		noisy	& & 2.29 (1.92) & \\\midrule
		MSE		& 3.16 (9.57)		& 3.10 (9.58)	& 3.17 (9.58 )\\
		MAE		& \textbf{3.25}	(\textbf{9.73}) 	& 3.08 (\textbf{9.68})	& \textbf{3.25} (\textbf{9.75}) \\
		LSD		& 3.04 (8.59)	& 3.03 (8.31)	& -- \\
		wLSD 	& 3.19 (9.12)	& \textbf{3.21} (8.88)	& -- \\
		Comp 	& \textbf{3.25} (9.45)	& 2.88 (9.21)	& \textbf{3.31} (9.42) \\
		SNR / SDR	& 3.15 (9.54)	& 3.11 (9.62)	& 3.19 (\textbf{9.66}) \\
		Corr 	& 3.16 (9.56)	& 3.11 (9.60)	& 3.16 (9.58) \\
		SDW		& 3.12 (9.61)	& --	& -- \\
		MSE-AMR	& 3.01 (9.39) 	& 2.98 (9.45) & -- \\
	\end{tabular}
\end{table}
While the PLSD shows no advantage over the \ac{LSD}, the wPLSD gives a slight advantage over the magnitude-based wLSD, which confirms the importance of attributing low weights to unimportant frequency bins for the \ac{LSD}.
It is not surprising that the \ac{SNR} and \ac{SDR} perform on par with the L2 norm, as they are merely normalized versions. The correlation-based losses are in the same range as well. 
It is surprising that on this reverberant dataset, the \ac{SDW} loss performs significantly worse than the magMSE or cMSE, which has been shown differently on non-reverberant datasets in \cite{Xia2020,Xu2020}. This highlights also the data dependency of the speech distortion weight $\lambda$, which varies from 0.3 in \cite{Xia2020}, 0.5 in \cite{Xu2020}, and 0.6 in our case. Furthermore, on the reverberant Chime2 dataset, we also could not confirm the effectiveness of perceptually motivated weightings, such as the AMR weighting proposed in \cite{Zhao2019}, which performed significantly worse than the unweighted MSEs.
While the \ac{SI-SDR} is less correlated with speech quality than \ac{PESQ}, it shows the best results mostly for linear losses such as \ac{MAE} and \ac{SDR}.
Overall we can say that magnitude compression and carefully chosen distance metrics according to the spectral domain's distribution can lead to more suitable loss functions.

\section{Conclusions}
We have classified several signal-based frequency domain loss functions for speech enhancement and exploited relations and performance differences on the reverberant Chime2 dataset. Our experiments showed that for such realistic data, compressed losses are beneficial and that combined magnitude and complex losses improve the objective speech quality. We also showed different findings for weighted losses with reverberant speech than for anechoic data. Future work has to be done especially on improved phase-aware losses to further improve the quality.


\bibliographystyle{IEEEbib}
\bibliography{shortened_bib,sapref.bib}

\begin{thebibliography}{10}

\bibitem{Wang2018}
D.~{Wang} and J.~{Chen},
\newblock ``Supervised speech separation based on deep learning: An overview,''
\newblock vol. 26, no. 10, Oct 2018.

\bibitem{Kavalekalam2018}
M.~S. {Kavalekalam}, J.~K. {Nielsen}, M.~G. {Christensen}, and J.~B. {Boldt},
\newblock ``A study of noise {PSD} estimators for single channel speech
  enhancement,''
\newblock in {\em Proc. {IEEE} Intl. Conf. on Acoustics, Speech and Signal
  Processing (ICASSP)}, 2018, pp. 5464--5468.

\bibitem{Williamson2017}
D.~S. {Williamson} and D.~{Wang},
\newblock ``Time-frequency masking in the complex domain for speech
  dereverberation and denoising,''
\newblock {\em {IEEE/ACM} Trans. Audio, Speech, Lang. Process.}, vol. 25, no.
  7, pp. 1492--1501, July 2017.

\bibitem{Ephrat2018}
A.~Ephrat, I.~Mosseri, O.~Lang, T.~Dekel, K.~Wilson, A.~Hassidim, W.~T.
  Freeman, and M.~Rubinstein,
\newblock ``Looking to listen at the cocktail party: A speaker-independent
  audio-visual model for speech separation,''
\newblock {\em ACM Trans. Graph.}, vol. 37, no. 4, July 2018.

\bibitem{Luo2019}
Y.~{Luo} and N.~{Mesgarani},
\newblock ``{Conv-TasNet}: Surpassing ideal time–frequency magnitude masking
  for speech separation,''
\newblock {\em {IEEE/ACM} Trans. Audio, Speech, Lang. Process.}, vol. 27, no.
  8, pp. 1256--1266, Aug 2019.

\bibitem{Strake2019}
M.~{Strake}, B.~{Defraene}, K.~{Fluyt}, W.~{Tirry}, and T.~{Fingscheidt},
\newblock ``Separated noise suppression and speech restoration: {LSTM}-based
  speech enhancement in two stages,''
\newblock in {\em WASPAA}, Oct 2019.

\bibitem{Wichern2017}
G.~{Wichern} and A.~{Lukin},
\newblock ``Low-latency approximation of bidirectional recurrent networks for
  speech denoising,''
\newblock in {\em Proc. {IEEE} Workshop on Applications of Signal Processing to
  Audio and Acoustics ({WASPAA})}, Oct 2017, pp. 66--70.

\bibitem{Wisdom2019}
S.~{Wisdom}, J.~R. {Hershey}, K.~{Wilson}, J.~{Thorpe}, M.~{Chinen},
  B.~{Patton}, and R.~A. {Saurous},
\newblock ``Differentiable consistency constraints for improved deep speech
  enhancement,''
\newblock in {\em Proc. {IEEE} Intl. Conf. on Acoustics, Speech and Signal
  Processing (ICASSP)}, May 2019, pp. 900--904.

\bibitem{Tan2018}
K.~Tan and D.~Wang,
\newblock ``A convolutional recurrent neural network for real-time speech
  enhancement,''
\newblock in {\em Proc. Interspeech}, 2018, pp. 3229--3233.

\bibitem{Wang2014}
Y.~{Wang}, A.~{Narayanan}, and D.~{Wang},
\newblock ``On training targets for supervised speech separation,''
\newblock {\em {IEEE} Trans. Audio, Speech, Lang. Process.}, vol. 22, no. 12,
  pp. 1849--1858, Dec 2014.

\bibitem{Kolbaek2017}
M.~{Kolbæk}, D.~{Yu}, Z.~{Tan}, and J.~{Jensen},
\newblock ``Multitalker speech separation with utterance-level permutation
  invariant training of deep recurrent neural networks,''
\newblock vol. 25, no. 10, Oct 2017.

\bibitem{Bahmaninezhad2019}
F.~Bahmaninezhad, J.~Wu, R.~Gu, S.-X. Zhang, Y.~Xu, M.~Yu, and D.~Yu,
\newblock ``{A Comprehensive Study of Speech Separation: Spectrogram vs
  Waveform Separation},''
\newblock in {\em Proc. Interspeech 2019}, 2019, pp. 4574--4578.

\bibitem{Martin-Donas2018}
J.~M. {Martín-Doñas}, A.~M. {Gomez}, J.~A. {Gonzalez}, and A.~M. {Peinado},
\newblock ``A deep learning loss function based on the perceptual evaluation of
  the speech quality,''
\newblock {\em {IEEE} Signal Process. Lett.}, vol. 25, no. 11, pp. 1680--1684,
  Nov 2018.

\bibitem{Zhao2019}
Z.~{Zhao}, S.~{Elshamy}, and T.~{Fingscheidt},
\newblock ``A perceptual weighting filter loss for {DNN} training in speech
  enhancement,''
\newblock in {\em Proc. {IEEE} Workshop on Applications of Signal Processing to
  Audio and Acoustics ({WASPAA})}, Oct 2019, pp. 229--233.

\bibitem{Liu2017}
Q.~{Liu}, W.~{Wang}, P.~J.~B. {Jackson}, and Y.~{Tang},
\newblock ``A perceptually-weighted deep neural network for monaural speech
  enhancement in various background noise conditions,''
\newblock in {\em Proc. European Signal Processing Conf. (EUSIPCO)}, Aug 2017,
  pp. 1270--1274.

\bibitem{Kolbaek2020}
M.~{Kolbæk}, Z.~{Tan}, S.~H. {Jensen}, and J.~{Jensen},
\newblock ``On loss functions for supervised monaural time-domain speech
  enhancement,''
\newblock {\em IEEE/ACM Transactions on Audio, Speech, and Language
  Processing}, vol. 28, pp. 825--838, 2020.

\bibitem{Xia2020}
R.~Xia, S.~Braun, C.~Reddy, H.~Dubey, R.~Cutler, and I.~Tashev,
\newblock ``Weighted speech distortion losses for neural-network-based
  real-time speech enhancement,''
\newblock in {\em Proc. {IEEE} Intl. Conf. on Acoustics, Speech and Signal
  Processing (ICASSP)}, 2020.

\bibitem{Weninger2015}
F.~Weninger, H.~Erdogan, S.~Watanabe, E.~Vincent, J.~{Le Roux}, J.~R. Hershey,
  and B.~Schuller,
\newblock ``Speech enhancement with {LSTM} recurrent neural networks and its
  application to noise-robust {ASR},''
\newblock in {\em Proc. Latent Variable Analysis and Signal Separation}, 2015,
  pp. 91--99.

\bibitem{Zhao2018a}
H.~{Zhao}, S.~{Zarar}, I.~{Tashev}, and C.~{Lee},
\newblock ``Convolutional-recurrent neural networks for speech enhancement,''
\newblock in {\em Proc. {IEEE} Intl. Conf. on Acoustics, Speech and Signal
  Processing (ICASSP)}, 2018, pp. 2401--2405.

\bibitem{Fu2017}
S.~{Fu}, T.~{Hu}, Y.~{Tsao}, and X.~{Lu},
\newblock ``Complex spectrogram enhancement by convolutional neural network
  with multi-metrics learning,''
\newblock in {\em IEEE Intl. Workshop on Machine Learning for Signal Processing
  (MLSP)}, 2017, pp. 1--6.

\bibitem{Wang2020}
Z.~{Wang} and D.~{Wang},
\newblock ``Multi-microphone complex spectral mapping for speech
  dereverberation,''
\newblock 2020.

\bibitem{Ephraim1985}
Y.~Ephraim and D.~Malah,
\newblock ``Speech enhancement using a minimum mean-square error log-spectral
  amplitude estimator,''
\newblock vol. 33, no. 2, 1985.

\bibitem{Tu2018}
Y.-H. {Tu}, I.~{Tashev}, S.~{Zarar}, and C.~{Lee},
\newblock ``A hybrid approach to combining conventional and deep learning
  techniques for single-channel speech enhancement and recognition,''
\newblock in {\em Proc. {IEEE} Intl. Conf. on Acoustics, Speech and Signal
  Processing (ICASSP)}, April 2018, pp. 2531--2535.

\bibitem{Lee2018}
J.~{Lee}, J.~{Skoglund}, T.~{Shabestary}, and H.~{Kang},
\newblock ``Phase-sensitive joint learning algorithms for deep learning-based
  speech enhancement,''
\newblock {\em IEEE Signal Processing Letters}, vol. 25, no. 8, pp. 1276--1280,
  2018.

\bibitem{Wilson2018}
K.~{Wilson}, M.~{Chinen}, J.~{Thorpe}, B.~{Patton}, J.~{Hershey}, R.~A.
  {Saurous}, J.~{Skoglund}, and R.~F. {Lyon},
\newblock ``Exploring tradeoffs in models for low-latency speech enhancement,''
\newblock in {\em IWAENC}, Sep. 2018, pp. 366--370.

\bibitem{Vincent2006a}
E.~Vincent, R.~Gribonval, and C.~Fevotte,
\newblock ``Performance measurement in blind audio source separation,''
\newblock vol. 14, no. 4, July 2006.

\bibitem{Choi2019}
H.-S. Choi, J.~Kim, J.~Hur, A.~Kim, J.-W. Ha, and K.~Lee.,
\newblock ``Phase-aware speech enhancement with deep complex {U-Net},''
\newblock in {\em Intl. Conf. on Learning Representations (ICLR)}, 2019.

\bibitem{Roux2019}
J.~L. {Roux}, S.~{Wisdom}, H.~{Erdogan}, and J.~R. {Hershey},
\newblock ``{SDR} - half-baked or well done?,''
\newblock May 2019.

\bibitem{Heitkaemper2020}
J.~{Heitkaemper}, D.~{Jakobeit}, C.~{Boeddeker}, L.~{Drude}, and
  R.~{Haeb-Umbach},
\newblock ``Demystifying tasnet: A dissecting approach,''
\newblock in {\em Proc. {IEEE} Intl. Conf. on Acoustics, Speech and Signal
  Processing (ICASSP)}, 2020, pp. 6359--6363.

\bibitem{Taal2011a}
C.~H. Taal, R.~C. Hendriks, R.~Heusdens, and J.~Jensen,
\newblock ``An algorithm for intelligibility prediction of time-frequency
  weighted noisy speech,''
\newblock {\em {IEEE} Trans. Audio, Speech, Lang. Process.}, vol. 19, no. 7,
  pp. 2125--2136, Sept 2011.

\bibitem{Venkataramani2017}
S.~Venkataramani, J.~Casebeer, and P.~Smaragdis,
\newblock ``Adaptive front-ends for end-to-end source separation,''
\newblock in {\em Conference on Neural Information Processing Systems (NIPS)},
  2017.

\bibitem{Xu2020}
Z.~{Xu}, S.~{Elshamy}, and T.~{Fingscheidt},
\newblock ``Using separate losses for speech and noise in mask-based speech
  enhancement,''
\newblock in {\em Proc. {IEEE} Intl. Conf. on Acoustics, Speech and Signal
  Processing (ICASSP)}, 2020, pp. 7519--7523.

\bibitem{Cho2014}
K.~Cho, B.~Van Merri{\"e}nboer, D.~Bahdanau, , and Y.~Bengio,
\newblock ``On the properties of neural machine translation: Encoder-decoder
  approaches,''
\newblock in {\em Proc. 8th Workshop on Syntax, Semantics and Structure in
  Statistical Translation (SSST-8)}, 2014.

\bibitem{Loshchilov2019}
Ilya Loshchilov and Frank Hutter,
\newblock ``Decoupled weight decay regularization,''
\newblock in {\em International Conference on Learning Representations}, 2019.

\bibitem{ITU_T_P862}
ITU-T,
\newblock ``Perceptual evaluation of speech quality ({PESQ}), an objective
  method for end-to-end speech quality assessment of narrowband telephone
  networks and speech codecs,'' Feb. 2001.

\bibitem{Vincent2013}
E.~Vincent, J.~Barker, S.~Watanabe, and F.~Nesta,
\newblock ``The second '{CHIME}' speech separation and recognition challenge:
  datadata, tasks and baselines,''
\newblock in {\em Proc. {IEEE} Intl. Conf. on Acoustics, Speech and Signal
  Processing (ICASSP)}, June 2012.

\end{thebibliography}

\end{document}